\documentclass[journal=cmatex,manuscript=article]{achemso}

\usepackage{soul}
\usepackage{chemformula} 
\usepackage[T1]{fontenc} 

\author{Matthias Krinninger}
	\affiliation[TUMPC]
	{Technical University of Munich, TUM School of Natural Sciences, Department of Chemistry, Chair of Physical Chemistry, Lichtenbergstr.~4,D-85748 Garching, Germany}
	\alsoaffiliation[CRC]{Catalysis Research Center, Technical University of Munich, Ernst-Otto-Fischer-Str.~1, D-85748 Garching, Germany}
	\alsoaffiliation[TUMFNG]
	{Technical University of Munich, TUM School of Natural Sciences, Department of Chemistry, Functional Nanomaterials Group, Lichtenbergstr.~4, D-85748 Garching, Germany}
\author{Nicolas Bock}
	\affiliation[TUMPC]
	{Technical University of Munich, TUM School of Natural Sciences, Department of Chemistry, Chair of Physical Chemistry, Lichtenbergstr.~4,D-85748 Garching, Germany}
	\alsoaffiliation[CRC]{Catalysis Research Center, Technical University of Munich, Ernst-Otto-Fischer-Str.~1, D-85748 Garching, Germany}
\author{Sebastian Kaiser}
	\affiliation[TUMPC]
	{Technical University of Munich, TUM School of Natural Sciences, Department of Chemistry, Chair of Physical Chemistry, Lichtenbergstr.~4,D-85748 Garching, Germany}
	\alsoaffiliation[CRC]{Catalysis Research Center, Technical University of Munich, Ernst-Otto-Fischer-Str.~1, D-85748 Garching, Germany}
\author{Johanna Plansky}
	\affiliation[TUMFNG]
	{Technical University of Munich, TUM School of Natural Sciences, Department of Chemistry, Functional Nanomaterials Group, Lichtenbergstr.~4, D-85748 Garching, Germany}
	\alsoaffiliation[CRC]{Catalysis Research Center, Technical University of Munich, Ernst-Otto-Fischer-Str.~1, D-85748 Garching, Germany}
\author{Tobias Bruhm}
	\affiliation[TUMCat]
	{Technical University of Munich, TUM School of Natural Sciences, Department of Chemistry, Professorship of Inorganic Chemistry, Lichtenbergstr.~4, D-85748 Garching, Germany}
	\alsoaffiliation[CRC]{Catalysis Research Center, Technical University of Munich, Ernst-Otto-Fischer-Str.~1, D-85748 Garching, Germany}
\author{Felix Haag}
\author{Francesco Allegretti}
	\affiliation[TUME20]
	{Technical University of Munich, TUM School of Natural Sciences, Department of Physics, Chair of Experimental Physics (E20), James-Franck Str.~1, D-85748 Garching, Germany}
\author{Ueli Heiz}
	\affiliation[TUMPC]
	{Technical University of Munich, TUM School of Natural Sciences, Department of Chemistry, Chair of Physical Chemistry, Lichtenbergstr.~4,D-85748 Garching, Germany}
	\alsoaffiliation[CRC]{Catalysis Research Center, Technical University of Munich, Ernst-Otto-Fischer-Str.~1, D-85748 Garching, Germany}
\author{Klaus K\"ohler}
	\affiliation[TUMCat]
	{Technical University of Munich, TUM School of Natural Sciences, Department of Chemistry, Professorship of Inorganic Chemistry, Lichtenbergstr.~4, D-85748 Garching, Germany}
	\alsoaffiliation[CRC]{Catalysis Research Center, Technical University of Munich, Ernst-Otto-Fischer-Str.~1, D-85748 Garching, Germany}
\author{Barbara A.J. Lechner}
	\affiliation[TUMFNG]
	{Technical University of Munich, TUM School of Natural Sciences, Department of Chemistry, Functional Nanomaterials Group, Lichtenbergstr.~4, D-85748 Garching, Germany}
	\alsoaffiliation[CRC]{Catalysis Research Center, Technical University of Munich, Ernst-Otto-Fischer-Str.~1, D-85748 Garching, Germany}
	\alsoaffiliation[IAS]{Institute for Advanced Study, Technical University of Munich, Lichtenbergstr.~2a, D-85748 Garching, Germany}
\author{Friedrich Esch}
	\email{friedrich.esch@tum.de}
	\affiliation[TUMPC]
	{Technical University of Munich, TUM School of Natural Sciences, Department of Chemistry, Chair of Physical Chemistry, Lichtenbergstr.~4,D-85748 Garching, Germany}
	\alsoaffiliation[CRC]{Catalysis Research Center, Technical University of Munich, Ernst-Otto-Fischer-Str.~1, D-85748 Garching, Germany}

\title[An \textsf{achemso} demo]
{
	On-Surface Carbon Nitride Growth from Polymerization of 2,5,8-Triazido-\textit{s}-heptazine  
}

\abbreviations{IR,NMR,UV,UHV,TPD,TPR,STM,nc-AFM,TAH,HOPG,QMS}

\begin{document}

\begin{tocentry}
\begin{center}
\includegraphics[height=3.5cm]{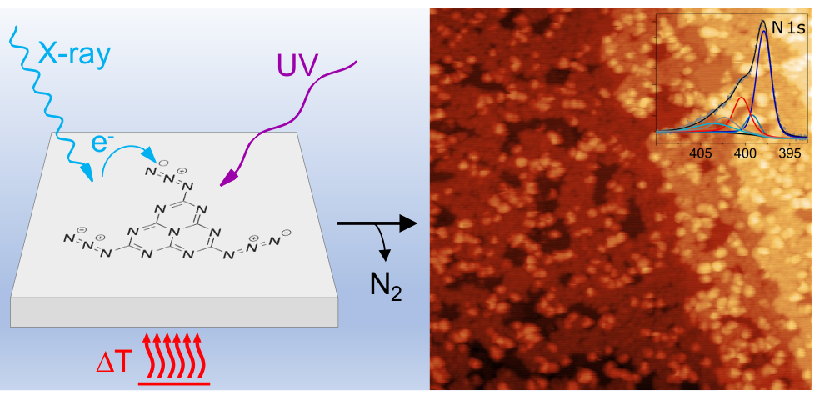}
\end{center}
\end{tocentry}
\newpage
\begin{abstract}
  Carbon nitrides have recently come into focus for photo- and thermal catalysis, both as support materials for metal nanoparticles as well as photocatalysts themselves. While many approaches for the synthesis of three-dimensional carbon nitride materials are available, only top-down approaches by exfoliation of powders lead to thin film flakes of this inherently two-dimensional material. Here, we describe an \textit{in situ} on-surface synthesis of monolayer 2D carbon nitride films, as a first step towards precise combination with other 2D materials. Starting with a single monomer precursor, we show that 2,5,8-triazido-\textit{s}-heptazine (TAH) can be evaporated intact, deposited on a single crystalline Au(111) or graphite support, and activated via azide decomposition and subsequent coupling to form a covalent polyheptazine network. We demonstrate that the activation can occur in three pathways, via electrons (X-ray illumination), photons (UV illumination) and thermally. Our work paves the way to coat materials with extended carbon nitride networks which are, as we show, stable under ambient conditions. 
\end{abstract}

\section{Introduction}
The long-established carbon nitrides, consisting of the earth-abundant elements carbon and nitrogen, have recently moved into focus of research for their potential as (photo)catalysts in water splitting or carbon dioxide photoreduction \cite{wang2012polymeric,kumar2019c3n5,schlomberg2019structural}. The term carbon nitride generally describes polymers consisting predominantly of C and N species. More specifically, (graphitic) carbon nitrides (g-\ch{C3N4}) are categorized as 1D or 2D networks based on the \textit{s}-triazine and \textit{s}-heptazine building blocks, in line with Lau and Lotsch. \cite{lau2022tour} Networks consisting of heptazine building blocks can be connected via tertiary amine-links, forming true g-\ch{C3N4} sheets where nitrogen atoms provide a coordination environment in the pores. Alternatively, connection via secondary amines leads to the formation of hydrogen-containing chains (melon) or cyclic arrangements in 2D poly(heptazine imides). Beyond the ideal structure of g-\ch{C3N4}, defective modifications have been designed to tune the electronic properties \cite{wang2009metal,yu2021point,lau2022tour}.

Besides their direct application for metal-free catalysis \cite{wang2009metal,li2019strategy}, the porous nanostructures thus provide binding sites for catalytically active metal ions or particles. The coordination of metals to the nitrogen atoms of the carbon nitride pores is flexible enough to provide access for reactants while also immobilizing the active sites efficiently against dissolution and sintering. As the group of Ramirez showed, Suzuki coupling can be sustainably performed even on single Pd ions that remain coordinated in the carbon nitride pores throughout the involved redox process \cite{chen2018heterogeneous}. Furthermore, the addition of metal and metal oxide cocatalysts into the pores has been shown to overcome sluggish charge transfer kinetics in the photo(electro)catalytic water splitting by carbon nitrides \cite{lau2022tour}. More generally, the ability to bind catalyst particles from single atoms to entire clusters in an ordered manner \cite{hosseini2019adsorption} could provide a platform to systematically access size effects in the catalysis on small clusters at the non-scalable limit \cite{crampton2017can, fukamori2013fundamental}.

Synthesis of purely 2D-linked, hydrogen-free g-\ch{C3N4} networks is experimentally highly challenging \cite{lau2022tour}. Starting from a variety of precursors \cite{kessler2017functional}, the thermal condensation of melamine typically leads to the formation of only 1D melon chains, while a guided synthesis mediated by salt melts or on crystalline salt surfaces induces the 2D order via formation of poly(heptazine imides) without terminating primary amines \cite{lau2022tour, guo2021surface}.
A variety of more complex approaches have been reported towards a 2D preparation of g-\ch{C3N4}, namely laser-electric discharge methods \cite{burdina2000synthesis}, chemical vapor deposition \cite{kouvetakis1994novel}, electrodeposition \cite{li2004graphitic}, spray deposition \cite{sima2019graphitic}, liquid-gas interface separation \cite{algara2014triazine}, and exfoliation \cite{bojdys2013exfoliation}. Furthermore, in an alternative synthetic approach, Gillan and later the Kroke groups demonstrated that extended networks can be formed starting from single triazine or heptazine azide precursors \cite{gillan2000synthesis, miller2004synthesis,miller2007nitrogen,Saplinova2009,Schwarzer2013}. Here, via thermal activation, without the need for a catalyst, powders of carbon nitride networks form. This route proceeds in the solid state via the intermediate formation of a highly reactive nitrene species; upon reaction, the sp\textsuperscript{2} ring character is maintained. The authors showed that the final hydrogen content and thus the degree of interlinking strongly depend on pressure conditions, humidity, exposure time and especially on heating rate upon thermal activation. This autocatalytic one-pot reaction is highly exothermic and occurs in a temperature range where the background slope of the thermogravimetric analysis might be interpreted as onset of evaporation. The various products that can be envisaged are summarized in the Supporting Information Fig.~S1 - from primary, secondary and tertiary amines to azo coupling, introducing hydrogen to the reactive nitrene sites of the network that is picked up from the ambient. Interestingly, the authors discuss only amine linking as final connection motif.

The challenge is now to bring the polymerization reactions known for powders of carbon nitride networks onto a highly defined 2D support in order to obtain single layer carbon nitrides as controllable interfaces. Since the formation of single atomic layers is still hardly accessible with current methods \cite{jiang2014dependence}, we chose the catalyst-free Gillan approach as the most promising one for translation of a powder to an on-surface synthesis. In this work, we present an azide-based polymerization route on two highly defined, atomically flat materials, i.e. Au(111) and highly oriented pyrolytic graphite (HOPG) that have the potential to be used as model electrodes for more applied future studies. As the single precursor, we opted for the hydrogen-free 2,5,8-triazido-\textit{s}-heptazine (TAH, see Fig.~S1) where three \ch{-N3} azide groups can react both thermally \cite{miller2004synthesis,miller2007nitrogen} and light-induced \cite{zheng2004theoretical, l1969decomposition} and, as we will show, also induced by electrons. For similar azides, a tradeoff between volatility and reactivity has been demonstrated to allow intact sublimation and deposition \cite{diaz2013surface,hellerstedt2019aromatic}, maintaining the azide reactivity, for example for a "click" reaction \cite{diaz2013surface}. The TAH precursor forms layered, close-packed crystals that indicate the feasibility of co-planar adsorption and flat 2D coupling on the surface \cite{miller2004synthesis}. The decomposition, under formation of gaseous nitrogen as only byproduct, leads to the formation of planar, rigid building blocks with directional, active nitrene linkers and represents therefore one of three reaction stimuli discussed to prepare ordered covalent organic frameworks (COFs, via the "single reaction pathway" in ref.\cite{haase2020solving}).

We base the discussion of our results on two recent ultrahigh vacuum (UHV) studies on on-surface reactions of similarly functionalized molecules: A first study on the thermally induced reactivity of single azide-functionalized phenanthrenes by highly resolving, cryogenic non-contact atomic force microscopy (nc-AFM) could discern three different reaction channels on Ag(111). Starting from a postulated silver-nitrenoid intermediate, the authors observed either the nitrene insertion into a \ch{C-H} bond, its dimerisation, and hydrogenation \cite{hellerstedt2019aromatic}. In comparison, our choice of the TAH precursor is free of hydrogen, allowing us to exclude the \ch{C-H} insertion pathway - that would lead to heptazine ring modifications - and thus to expect linking predominantly via azo bonds. In a second, rare example of surface-bound azide photochemistry, Luo et al. postulate for 4-methoxyphenyl azide on Cu(100) a mechanism that proceeds via a upright arylnitrene-intermediate where the reactive nitrene is stabilized by a bond to the support before coupling amongst the molecules to form a planar azoarene \cite{luo2019photochemical}.

In this paper, we focus on the various non-catalytic methods to activate the TAH azide groups, namely thermally, by photons and by electrons, and characterize the resulting films with a particular view on density, flatness, homogeneity and composition. Furthermore we show the stability of the resulting film in air, thus presenting a universal approach to anchoring catalytically active particles in highly controlled interfaces of porous carbon nitride networks on supports or in 2D-layered heterostructures \cite{yan2017tuning,bock2021towards}.

\section{Experimental}
The carbon nitride precursor 2,5,8-triazido-\textit{s}-heptazine (TAH) was synthesized from melamine over melem and 2,5,8-trihydrazino-\textit{s}-heptazine\cite{sattler2010investigations,Saplinova2009}.
Au(111) single crystal samples were prepared by several cycles of sputtering (Ar\textsuperscript{+}, 4$\times$10\textsuperscript{-5}~mbar, 2.0~keV, 15~min) and subsequent annealing (900~K, 10~min). For heating, a boron nitride heater located in the sample holder was used. The temperature was measured via a thermocouple (type~K) attached to the crystal.
For scanning tunneling (STM) measurements in air we used Au(111) laminated on mica samples (Georg Albert PVD-Beschichtungen) which were flame annealed in a hydrogen flame for several seconds.
Evaporation of TAH was performed with a home-built molecular evaporator consisting of a small tantalum crucible spot-welded on a wire connected to an electrical feedthrough. Additionally, a thermocouple (type~K) is spot-welded at the crucible that is heated via resistive heating.
STM measurements in ultra high vacuum (UHV, $p$\textsubscript{bg} < 1$\times$10\textsuperscript{-10} mbar) were performed with an Omicron VT-SPM in constant current mode with electrochemically etched tungsten tips at room temperature.
STM in air was performed with a STM built by the Wandelt group\cite{Wilms1999}, equipped with the SPM 100 control electronics by RHK. Here, tips were cut from a Pt/Ir wire (80/20, Temper hard, $d = 0.25$ mm, $R = 6.81~\Omega$m\textsuperscript{-1}).
For temperature programmed reaction (TPR) measurements, a so-called sniffer was used. A detailed description about this device can be found elsewhere\cite{kaiser2021cluster}. In short, desorption and reaction products are guided through a quartz tube, which is in close vicinity to the sample surface ($\sim$100~--~200 $\mu$m), to a differentially pumped quadrupole mass spectrometer (QMS, Pfeiffer Vacuum GmbH, QMA 200 Prisma Plus).
X-ray photoelectron spectroscopy measurements were performed with the Al~K$\alpha$-line of a SPECS XR~50 X-ray source and a Omicron EA~125 energy analyzer.

\section{Results and discussion}
Prerequisite for the clean preparation of polyheptazine films from an azide precursor is the feasibility of physical vapour deposition of the pure precursor, which was synthesized from melamine over melem and 2,5,8-trihydrazino-\textit{s}-heptazine \cite{sattler2010investigations,Saplinova2009}, as described in the Supporting Information, Section~S2; precursor purity and reactivity agree with literature \cite{miller2007nitrogen}. Hereby, two effects have to be balanced: While the high reactivity of the TAH azide functional groups calls for the lowest possible evaporation temperature, considerable intermolecular interactions within the TAH packing, due to the C/N alternation and charge distribution, imply rather high evaporation temperatures. We thus first had to determine whether we could evaporate the molecule, intact or already activated. To this purpose, a few milligram of the substance were heated in high vacuum ($p$\textsubscript{bg}<$5\times10^{-8}$~mbar) with the crucible placed in front of a quadrupole mass spectrometer (QMS), while monitoring the mass scan. As seen in Fig.~S3, a distinct peak of intact TAH (\ch{C6N16}, $m/z=296$) can be detected around 470~K, as well as several fragment signals in the range between $m/z=52$ and $m/z=132$, most likely originating from fragmentation in the QMS. We assign the fragment $m/z=78$ to \textit{s}-triazine (\ch{C3N3}), which has a comparatively strong intensity due to its stabilizing $\pi$-system. Further signals are present throughout the entire temperature range and arise from residual gases in the vacuum chamber, i.e. \ch{H2O} ($m/z=18$), CO ($m/z=28$) and \ch{CO2} ($m/z=44$). In the entire measurement range from $m/z=1-600$ we only observe signals from the monomer, its fragments, and residual gases.

Distinct spikes in the fragment signals around the TAH desorption maximum hint at possible molecule ejection caused by autocatalytic microexplosions at slightly hotter spots inside the crucible-located powder. Since both, the decomposition of TAH upon evaporation and the fragmentation of the evaporated TAH in the QMS, contribute to the atomic N signal ($m/z=14$), we discriminate the two by overlaying the TAH signal onto its initial rise: The polymerization reaction thus sets in at the temperature where the nitrogen curve starts to deviate from the TAH curve (indicated by a vertical dashed line in Fig.~S2b). We therefore deduce the ideal evaporation temperature range to be just below 445~K. 

In a next step, we dosed TAH onto a clean Au(111) surface and investigated three different pathways towards on-surface polymerization: (i) via X-ray illumination (and thus secondary electron excitation), (ii) via UV illumination and (iii) thermally. 

\textit{Electron-induced polymerization (via X-rays)}: We start with the polymerization induced by X-ray photons, as shown in Fig.~1. The network resulting from X-ray illumination at room temperature shows a random arrangement of bright protrusions of around 0.2~nm height from immobilized molecules, which confine similarly high streaks resulting from species that are mobile under the STM tip (Fig.~1a). The quality of the network can be improved and the streaks removed by heating to 573~K, a temperature that is sufficient to desorb monomers. The resulting network (Fig.~1b) appears more homogeneous in height, as clearly seen in the profiles in Fig.~1f. We find that the network still covers evenly the surface, without a net preference for nucleation at steps. When zooming in from Fig.~1b to c,d, we can resolve round features which in some places form a quasi-hexagonal network. From the measured average distance of about $0.9\pm0.1$ nm, approximately matching the molecule distances when azo-bridged ($0.8$ nm for free-standing dimers), we assign these features to individual heptazine molecules. The increased hole size upon further heating to 673~K in Fig.~1e suggests that the network contracts. Comparing height profiles shows that the corrugation within the polymer islands remains as flat as after the first annealing step.

\begin{figure}
	\centering
    \includegraphics[width=6.5in]{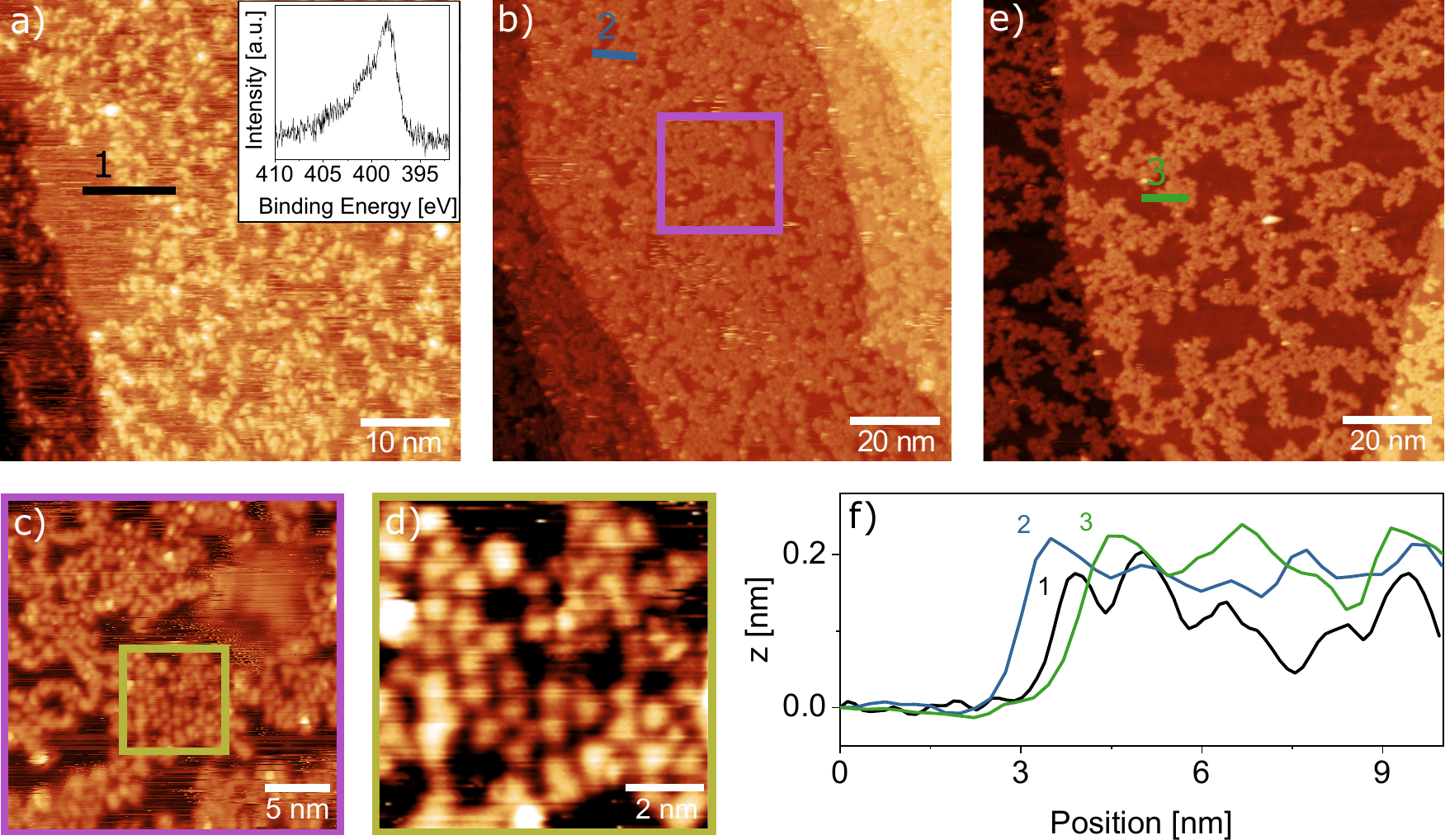}
	\caption
	{STM images of the on-surface synthesis of 2D carbon nitride via X-ray induced electron activation at RT. (a) TAH molecules evaporated onto Au(111) and subsequently illuminated by an X-ray source form a disordered network of molecules which covers most of the surface, with some streaky patches of mobile species. The inset shows an XPS measurement of the N~1s region, demonstrating that the azides have already been fully reacted off. (b) Annealing the same surface to 573~K leads to a less corrugated film and fewer streaks. (c,d) Zooming into the network reveals small areas of ordered molecules, which appear to remain intact upon polymerization. (e) A final annealing step to 673~K followed by further X-ray illumination opens up larger holes in the film. (f) Height profiles across the films in (a), (b) and (e) show that the overall height is similar in all three cases, corresponding to a monolayer. \textit{STM imaging parameters}: $U_b = 1.5$~V, (a-d) $I_t = 300$~pA, (e) $I_t = 10$~pA.
	}
	\label{fig:X-ray_film}
\end{figure}

To obtain further information on the chemical composition immediately after TAH deposition, we look at the N~1s XPS spectrum taken right after starting the X-ray illumination (Fig.~1a inset). As we will discuss in more detail below, this shape is characteristic for TAH molecules that have already lost their azide functional groups \cite{Zhang2019}. The same effect appears much slower on the less reactive HOPG surface, where we initially can record reference spectra of the intact precursor (a time-resolved XPS series on HOPG is shown in Section~S4). The N~1s spectrum in Fig.~1a shows that the azide reacts off quasi instantaneously upon illumination and only the heptazine core and the linking nitrogen species remain. This difference in reactivity suggests that secondary electrons, generated in the support, play a prominent role in X-ray activation. 

\textit{Photoinduced polymerization (UV light)}: Next, we look at the photoinduced synthesis by UV-illumination with an LED at 365~nm, as shown in Fig.~2. Turning the UV lamp on and off in front of a mass spectrometer shows the release of nitrogen only during illumination (Fig.~2a). To avoid competing thermal reaction channels, we performed this activation at 150~K and find that the reaction proceeds with an exponentially decaying rate that becomes negligible after a total illumination time of 10~min. The resulting film shown in Fig.~2b appears at first sight similarly holey to that in Fig.~1. However, height profiles across the structures show a film thickness that is up to twice as high and zooming in reveals fern-like details at different apparent heights. These parallel stripes at a distance of some~\AA~appear as though heptazine molecules standing upright are stacked in sheets. Such upright geometries have previously been reported on Cu(111) for nitrenoids \cite{luo2019photochemical} and for thermally activated melamine molecules \cite{lin2013selfassembly}, while the stacking recalls the layered 3D crystal structure (0.3~nm interlayer distance) \cite{miller2004synthesis}. This interpretation would suggest that we do not have a fully polymerized film at this stage. While we found that the as-deposited molecules are bound too weakly for stable STM imaging, here we now have an only partially reacted precursor state that we can image successfully, most likely thanks to direct nitrene-gold bonds. 
Indeed, a subsequent annealing step to 573~K leads to a flat, nearly complete single layer polymer with additional individual molecules or small agglomerates in the second layer. The height profile shown in Fig.~2d shows almost no corrugation within the monolayer areas of this annealed film.

\begin{figure}
	\centering
    \includegraphics[width=6.5in]{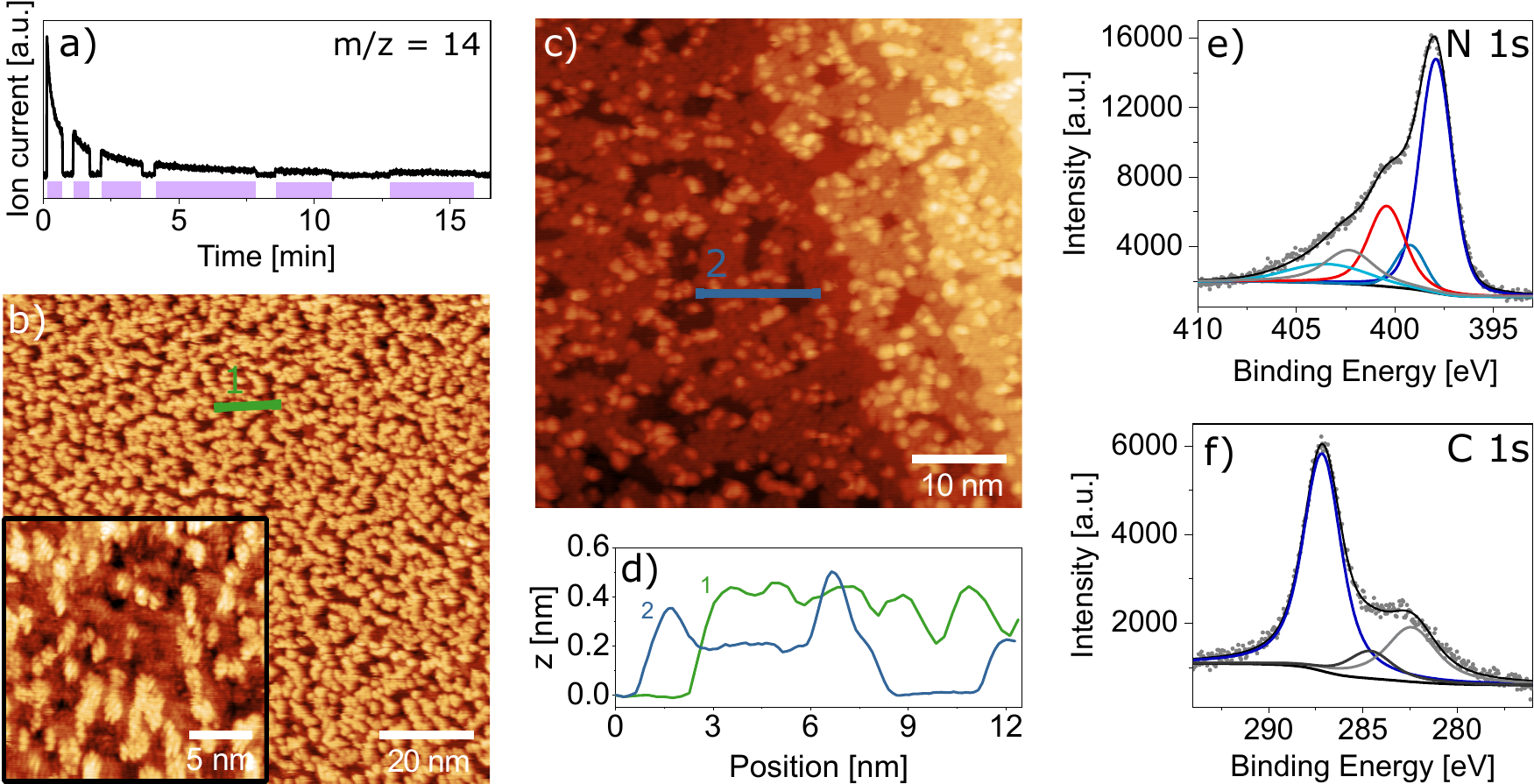}
	\caption
	{Room temperature STM images of the photoinduced synthesis of 2D carbon nitride at 150~K. (a) The nitrogen ($m/z=14$) mass trace demonstrates reproducible \ch{N2} release concomitantly to illumination with 365 nm UV light of TAH deposited onto Au(111). The purple shading indicates when the sample was illuminated. (b) STM images of the illuminated surface show an amorphous arrangement of molecules. The zoomed-in region in the inset reveals a fern-like appearance of molecule assemblies, which we interpret as stacked, upright-standing molecules. (c) Annealing the same surface to 573~K leads to a flat, nearly complete monolayer with individual molecules in the second layer. (d) Height profiles of the images in (b) and (c) show that the fern-like structure appears of similar height as the second layer molecules. (e) XPS data in the N~1s region of the preparation shown in (c) reveals the unchanged heptazine unit, represented by the conjugated heptazine peak (dark blue), the central tertiary amine (blue) and a red peak at the position of linking amines and azo components (for the XPS analysis see Section~S5). Characteristic peaks of the azide groups are absent. In analogy to the literature, we assign the light-blue peak to $\pi$-excitation, while the gray peak is an unidentified component. (f) The corresponding C~1s peak contains components of the heptazine (dark blue), adventitious carbon (black) and an unknown component (gray). \textit{STM imaging parameters}: $U_b = 1.5$~V, $I_t = 300~$pA.
	}
	\label{fig:photo}
\end{figure}

XPS measurements of the annealed film in the N~1s and C~1s regions indicate that the heptazine core remains unmodified by the polymerization reaction (blue peaks in Fig.~2e,f), while an additional smaller peak at 400.4~eV (red peak) indicates  and/or azo groups. These include bridging amine and azo linkers as well as chain terminating primary amines. Note that the hydrogenation of nitrenoid compounds to amines most likely occurs by reaction with residual hydrogen and water. The precise assignment based on literature (ref. \cite{Zhang2019} and references therein) and the relative peak intensities are given in Section~S5 (Table~S3).

\textit{Thermal polymerization}: In order to better understand the chemical reaction occurring during the polymerization, we performed highly sensitive temperature programmed reaction measurements using our sniffer setup \cite{bonanni2011experimental, kaiser2021cluster}. 
In the \ch{N2} temperature programmed desorption (TPD) spectrum, we observe four components.
We discuss the \ch{N2} TPD by comparing the overall \ch{N2} formation ($m/z=28$ trace) with molecular desorption (using the most intense accessible fragment $m/z=78$). We use HOPG as a reference for intact desorption (Fig.~3a). Here, we find that the $m/z=28$ and $78$ signals have precisely the same shape, implying that they arise from fragmentation of the intact, desorbed TAH molecule in the QMS. By overlaying the two traces, we determine a sensitivity multiplication factor of 3800. Notably, the onset of desorption just beyond room temperature implies that the interaction of TAH with the support is substantially weaker than within the bulk material.

\begin{figure}
	\centering
    \includegraphics[width=3.1in]{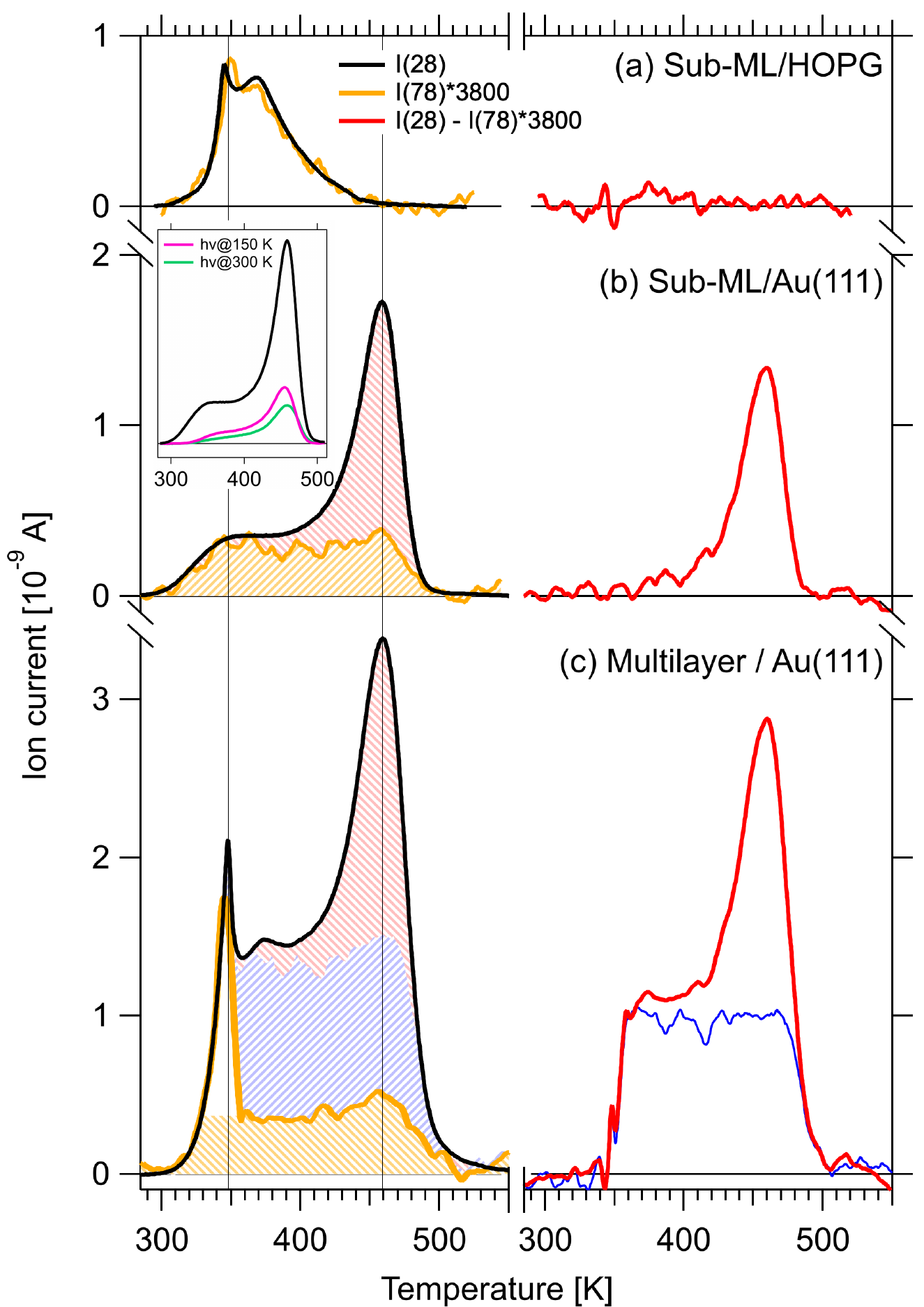}
	\caption
	{Temperature programmed desorption measurements during thermal on-surface carbon nitride synthesis. For each experiment, the mass traces for TAH molecules ($m/z=78$) and for \ch{N2} from azide decomposition ($m/z=28$) are shown in orange and black, respectively (left panels), the difference between the two traces in red (right panels). (a) By studying the desorption from HOPG, where no reaction but only multi- and monolayer desorption takes place, we calibrate the intensity difference to be a factor of 3800, which we apply to the $m/z=78$ traces on Au(111) in (b,c). (b) Desorption of a sub-ML coverage preparation on Au(111) shows two components, intact molecules desorbing in a broad peak between 300 and 500~K (orange) and reaction products from network formation around 450~K (red shaded area and difference trace). The inset compares the purely thermally treated sample with two that were previously illuminated for photoreaction at 150 (pink) and 300~K (green). Each spectrum starts from a similar TAH coverage. (c) At a higher TAH coverage, molecules desorb from multilayer sites around 350~K (white area in orange trace) and in the same broad peak again. Reaction products are now observed not only in the ~450~K peak but an additional broad component appears (the blue curve is a difference between the red curve in (c) and the red curve from (b) scaled by a factor of 1.4, resulting in the blue shaded area).}
	\label{fig:tpd}
\end{figure}

Applying the same scaling factor to a sub-monolayer sample on Au(111), shown in Fig.~3b, we observe that about 50~\% of the molecules desorb intact, while the remaining molecules react. To better illustrate the different components in the TPD curve, we show the difference spectra of the $m/z=28$ and the scaled $m/z=78$ signals in red in the right panels of Fig.~3. The difference curve for sub-ML/Au(111) shows a reactivity maximum around 450~K, which corresponds to the decomposition temperature of the pure powder precursor. In the $m/z=28$ curve (left), we can thus distinguish two components, intact desorption (shaded in orange) and reacted precursors (shaded in red), respectively. Notably, the orange component has a tabletop-like shape corresponding to a wide range of activation energies of desorption, and thus indicating a wide range of binding configurations (e.g. unspecific adsorption sites or varying intermolecular interactions). Once again the molecular desorption from the surface starts well below the temperature observed for powders.

When increasing the dosed TAH coverage to a multilayer equivalent, two more components appear in the TPD: A sharp peak at 350~K (white area in orange trace) is observed in both traces with intensities that match the scaling factor determined on HOPG. Note that the same peak is also apparent in the HOPG trace. We tentatively assign this peak to weakly, van der Waals-bound molecules in a highly disordered multilayer. The remaining yellow contribution is likely comparable to that of the sub-monolayer preparation, i.e. desorption of intact molecules. When subtracting both traces from each other, the same reaction peak at 450~K (red) is again observed, but with a slightly higher intensity and on a table-top background. To shade the different components in the multilayer TPD trace, we further subtracted the red curve in Fig.~3b, scaled by a factor of 1.4, from the red curve in Fig.~3c and obtain thus the fourth component, i.e. the blue curve and the blue shaded area, respectively. The blue component shows a similar behaviour as the yellow one and originates therefore most probably also from desorption of intact molecules. The reason for the missing signature in the $m/z=78$ trace remains unclear at this point.

From our TPD experiments we conclude that the polymerization reaction can indeed be quantitatively confirmed for up to 50~\% of the deposited molecules, by comparison of the peak ares of the different components. Surprisingly, the reaction temperature maximum coincides with the reaction onset for TAH in powder form (see Fig.~S1). The catalytic contribution of Au(111) to the polymerization reaction is thus minimal, suggesting that the film growth presented here is not support-specific and thus likely translatable. However, for purely thermal activation, the support must bind the TAH molecules sufficiently strongly to prevent complete desorption before the reaction temperature (as observed on HOPG). 

Such films formed by purely thermal synthesis are shown in Fig.~4. At very low coverages (Fig.~4a), small monolayer islands are formed that seem to follow the herringbone reconstruction by occupying preferentially fcc areas. Interestingly, at higher coverages (Fig.~4b), second layer molecules and agglomerates are observed before the first layer is completed, reflected in a highly corrugated line profile (Fig.~4c) similar to that observed for the photochemical activation followed by annealing. The XPS spectra in the N~1s and C~1s regions also appear similar to those in Fig.~2. The main peaks arising from heptazine ring and core (and their shake-up peak labelled "$\pi$-excitation" \cite{Zhang2019}), as well as combined side group signal of bridging and terminal nitrogen atoms, from here on called N~(side) atoms, correspond in their intensities (see Section~S5 and Table~S3 therein).

A direct, quantitative evaluation of the nitrogen species present in the N~1s spectra (Section~S5) can help to elucidate the carbon nitride stoichiometry and hence the type of network linking. In Table~S3, we calculate the ratio N~(side) atoms per heptazine unit as the ratio of the relative intensities of the N~(side) peak - that takes into account azo and amine linker species - to the heptazine core and ring N~1s signals, which together represent 7 N atoms or one heptazine unit. For both, the thermal and the photochemical+thermal preparations discussed in Figs.~2 and 4, we obtain similar values in the range of 2.0 -- 2.5 N~(side) atoms per heptazine unit. We can compare these values with the numbers expected for different linking motifs (Table~S2) and conclude that the obtained network contains more linking nitrogen atoms per heptazine unit than expected for a perfect 2D network formed by tertiary (1.0) or secondary amine nodes (1.5) - the linking motifs postulated e.g. by Gillan for TAH-derived carbon nitride powders \cite{miller2007nitrogen}. Azo groups, instead, and terminal amines have more N~(side) atoms per heptazine unit (3.0). Thus, we can as well exclude that our networks are exclusively azo-linked. The irreversible covalent bond formation upon polymerization favours rather amorphous and defect-rich networks with a high amount of terminal amines. Without their separate quantification, a clear attribution of the linking motifs remains elusive.

The \ch{N2} TPD traces resulting from these two preparation pathways (see Fig.~3b with inset) indicate that after the photoreaction - UV illumination at 150 (pink), resp. 300~K (green) - still a third of the azide groups remains intact. This intensity ratio supports the hypothesis that two of the three azide groups per molecule have reacted to a nitrene. If their reactivity is quenched by a direct nitrenoid bond to the gold support, this would point to an upright adsorption geometry of the molecules, as hypothesized for the striped patterns observed via STM in Fig.~2b and in analogy to the structures discussed in the literature by Luo et al. \cite{luo2019photochemical} and Lin et al.\cite{lin2013selfassembly} Furthermore, we note a slight difference in the shape of the TPD traces concerning the onset of the table-top desorption signal that occurs at slightly higher temperatures for the previously illuminated sample.

\begin{figure}
	\centering
    \includegraphics[width=6.5in]{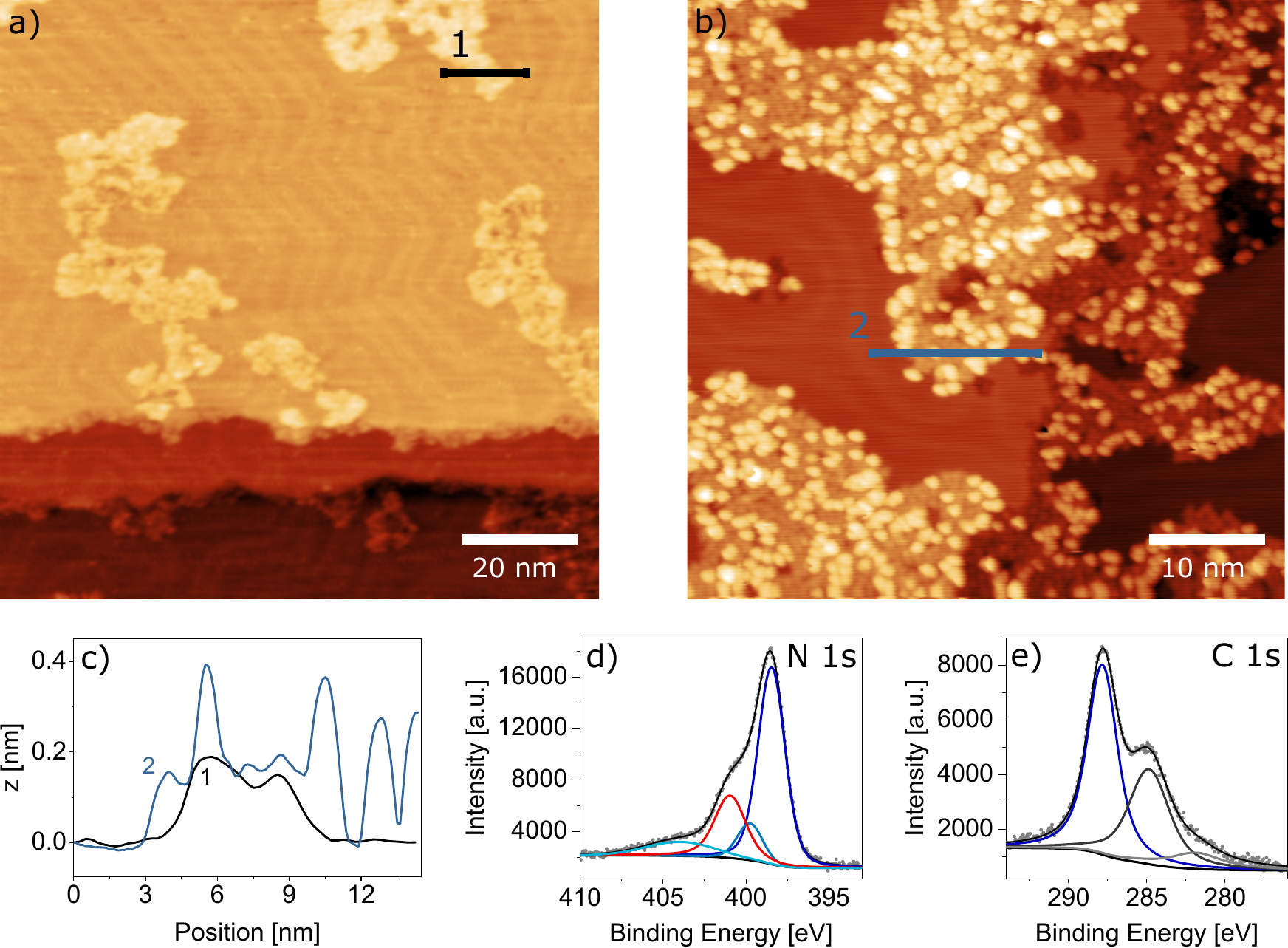}
	\caption
	{Thermal on-surface synthesis of carbon nitride films. (a) At a low coverage, it becomes apparent that the network grows in small islands which are constrained predominantly into the fcc areas of the Au(111) herringbone reconstruction. (b) At a higher coverage, the island area increases. Again, second layer molecules are observed and reveal an inherent tendency of the molecules to stack. Both STM images were recorded after thermal polymerization. (c) Height profiles from (a, b) show single layer islands for the low coverage and single layer islands with second layer molecules on top for the higher coverage. (d, e) XPS data in the N~1s and C~1s regions, respectively, show that the film formed by thermal reaction is comparable to that from photoreaction shown in Figure 2.
	Peak colors are assigned as before: Conjugated heptazine ring (dark blue), central tertiary amine (blue), $\pi$-excitation (light blue) and linking amines and azo components (red). The corresponding C~1s peak contains components of the heptazine (dark blue), adventitious carbon (black) and an unknown component (gray).
	\textit{STM imaging parameters}: $U_b = 1.5$~V, (a) $I_t = 300$~pA, (b) $I_t = 200$~pA.
	}
	\label{fig:therm}
\end{figure}

In a final step, we evaluated the stability of the carbon nitride film on Au(111) in air. To this purpose, the as-evaporated TAH film was transported through air and subsequently exposed to UV light in a tube furnace under vacuum. The resulting surface observed by ambient STM shown in Fig.~5a exhibits a remarkably flat film with a similarly holey network as the films prepared under UHV conditions. The N~1s spectrum taken after transferring back into a UHV chamber agrees well with the equivalent measurements in Figs.~1-4, but we refrain from fitting the spectrum due to poorer signal-to-noise ratio. The interpretation of the C~1s region is more difficult due to the presence of a large adventitious carbon peak from air-borne adsorbates. Nevertheless we can conclude that the obtained carbon nitride films are stable in air and can thus be used as supports for ambient pressure experiments.

\begin{figure}
	\centering
    \includegraphics[width=3.1in]{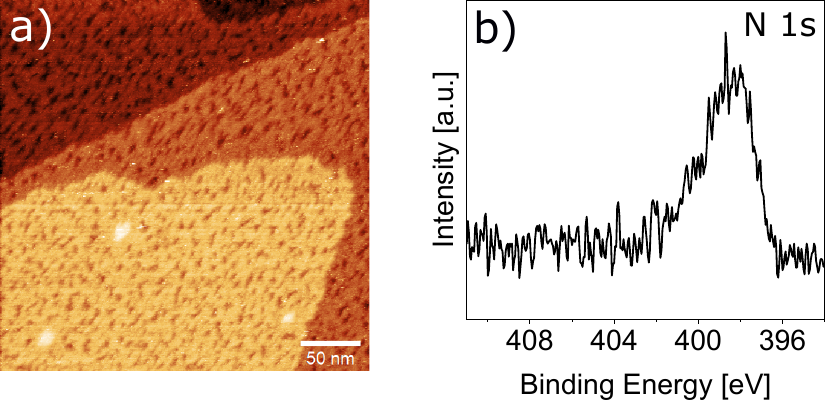}
	\caption
	{Stability of the carbon nitride film in air. (a) STM image of the film under ambient conditions (air, RT) after evaporation in a high vacuum setup, transport through air and subsequent illumination with UV light (385~nm) at RT in an evacuated quartz tube with a background pressure of $1\times10^{-2}$~mbar. \textit{STM imaging parameters}: $U_b = -215$~mV, $I_t = 200$~pA. (b) XPS N~1s region after equivalent preparation conditions.
	}
	\label{fig:air}
\end{figure}

\section{Conclusions}
In conclusion, we have demonstrated several pathways for on-surface synthesis of carbon nitride films starting from a single, azide-functionalized precursor with a heptazine core, namely 2,5,8-triazido-\textit{s}-heptazine (TAH). The simplicity of our approach lies in this particular, highly reactive precursor that has previously already been explored in powder chemistry. Here, we showed that we can successfully evaporate this relatively large molecule just before thermal activation of the azide occurs. We successfully polymerized the deposited TAH molecules by (a) X-ray illumination (and concomitant secondary electron emission), (b) UV-light illumination and subsequent annealing, and (c) purely thermal activation. All three pathways lead to a similar, somewhat holey, 2D film that is stable in UHV and ambient conditions. We demonstrate that the created amorphous film is largely homogeneous, with a stoichiometry pointing to various linking motifs. The UV-illuminated film provides an interesting insight into a partially reacted intermediate state before full polymerization by thermal annealing. We interpret the intermediates as upright standing nitrenoids with strong direct nitrene-gold bonds. While gold thus stabilizes intermediates, it does not have a catalytic role in the thermal polymerization reaction.

The presented synthesis route constitutes a highly versatile tool for preparing thin 2D carbon nitride films which can be included into stacked 2D materials or utilized as supports for nanoparticle experiments in vacuum, gas phase and likely also liquid environments. Similar preparations have been shown to successfully immobilize even single metal ions for catalytic applications. Here, we anticipate particularly their use as highly stable confining pores for supported nanoparticles in harsh (electro)catalytic environments. Specifically, we envisage the combination of the on-surface synthesis of 2D carbon nitride films with our recently demonstrated approach to deposit size-selected clusters from polyoxometalate precursors \textit{in situ} to be highly promising for atomic-scale studies on photo(electro)catalytic water splitting environments.

\begin{acknowledgement}

The authors thank Astrid de Clercq, Peter S. Deimel, Peter Feulner, and Johannes K\"uchle for experimental support and helpful discussions.
This work was funded by the Deutsche Forschungsgemeinschaft (DFG, German Research Foundation) under Germany's Excellence Strategy EXC 2089/1-390776260, through the project CRC1441 (project number 426888090), as well as by the grant ES 349/4-1 and TUM International Graduate School of Science and Engineering (IGSSE) via DFG, GSC~81. It received funding from the European Union's Horizon 2020 research and innovation programme under grant agreement no. 101007417 within the framework of the NFFA-Europe Pilot Joint Activities and under grant agreement no. 850764 from the European Research Council (ERC). B.A.J.L. gratefully acknowledges financial support from the Young Academy of the Bavarian Academy of Sciences and Humanities.

\end{acknowledgement}

\begin{suppinfo}

A pdf file containing the following information is available free of charge: S1. Reaction scheme of the polymerization reaction; S2. Description of the TAH synthesis and characterization of its purity; S3. Mass Spectra of the TAH evaporation process; S4. XPS and STM reference measurements on the TAH azide activation on HOPG, via X-ray-induced electron activation; S5. XPS analysis of the heptazine network on Au(111), as formed by the photo+thermal and the pure thermal reaction.

\end{suppinfo}

\bibliography{./library.bib}

\end{document}